\documentclass[prb,aps,twocolumn,showpacs,dvips]{revtex4}
\usepackage{feynmp}
\usepackage{amssymb}
\usepackage{epsfig}
\usepackage{graphicx}
\usepackage{subfigure}
\usepackage{mathrsfs}
\usepackage{hyperref}
\usepackage{cases}

\begin{document}

\title{Impact of ordering competition on the global phase diagram of iron pnictides}

\author{Jing Wang}
\affiliation{Department of Modern Physics, University of Science and
Technology of China, Hefei, Anhui 230026, P.R. China}
\author{Guo-Zhu Liu}
\affiliation{Department of Modern Physics, University of Science and
Technology of China, Hefei, Anhui 230026, P.R. China}

\begin{abstract}
We consider the impact of the competition among superconductivity,
spin density wave, and nematic order in iron pnictides, and show
that the ordering competition substantially reshapes the global
phase diagram. We perform a detailed renormalization group analysis
of an effective field theory of iron pnictides and derive the flow
equations of all the physical parameters. Using these results, we
find that superconductivity can be strongly suppressed by the
ordering competition, and also extract the $T$-dependence of
superfluid density. Moreover, the phase transitions may become first
order. Interestingly, our RG analysis reveal that the nematic order
exists only in an intermediate temperature region $T_{m}< T <
T_{n}$, but is destroyed at $T > T_{n}$ by thermal fluctuation and
at $T < T_{m}$ by ordering competition. This anomalous existence of
nematic order leads to a back-bending of the nematic transition line
on the phase diagram, consistent with the observed reentrance of
tetragonal structure at low temperatures. A modified phase diagram
is obtained based on the RG results.
\end{abstract}


\maketitle

%%%%%%%%%%%%%%%%%%%%%%%%%%%%%Main Body%%%%%%%%%%%%%%%%%%%%%%%%%%%%%%%%%%%%%

Competition between distinct long-range orders is a common
phenomenon that we frequently meet when studying a number of
unconventional superconductors, including cuprates
\cite{LeeRMP2006}, heavy fermion compounds \cite{Loehneysen,
Stockert}, and iron pnictides \cite{Kamihara2008JACP,
Chen2008Nature, Chen2008PRL, Rotter2008PRL, Fisher2011RPP,
Hirschfeld2011RPP, Chubukov2012}. Although ordering competition is a
general concept and occurs in various patterns, the most frequently
studied is that superconductivity compete and coexist with
antiferrogmagnetism or nematic order \cite{Vojta_rev}. When these
orders coexist in a bulk superconductor, one expects that a
well-defined quantum critical point (QCP) exists somewhere in the
superconducting (SC) dome. An important question is how to probe the
widely predicted QCP in realistic experiments.

Recently, we studied the physical effects of the competition between
superconductivity and nematic order in a $d$-wave cuprate
superconductor \cite{Liu2012PRB}, and found that the superfluid
density $\rho_s$ is suppressed at the nematic QCP significantly.
According to our analysis, the suppression of $\rho_s$ is indeed
caused by two scenarios. First, the ordering competition reduces the
charge condensate. Second, the gapless nodal quasiparticles couple
strongly to the critical fluctuation of nematic order, which excites
more normal quasiparticles out of the condensate. We further showed
that the suppression effect is significant solely at the QCP
\cite{Liu2012PRB}, thus $\rho_s$ should exhibit a deep valley at
this point. Based on these results, we proposed \cite{Liu2012PRB}
that the nematic QCP can be probed by measuring London penetration
depth $\lambda_{L}$, which satisfies $\lambda_{L}^{-2} \propto
\rho_s$. Clearly, the deep valley of $\rho_s$ corresponds to a sharp
peak of $\lambda_{L}$.

Thus far, no experimental evidence for the suppression of superfluid
density has been reported in cuprates. It is interesting that
Hashimoto \emph{et al.} have measured the penetration depth
$\lambda_{L}$ in an iron pnictide BaFe$_2$(As$_{1-x}$P$_x$)$_2$ and
observed a sharp peak of $\lambda_{L}$ \cite{Hashimoto2012Science},
which was claimed to signal the existence of a QCP of certain
competing order beneath the SC dome \cite{Hashimoto2012Science}.
This observation has stimulated considerable theoretical interest
\cite{Chowdhury2013PRL, Fernandes2013PRL, Levchenko2013PRL} on the
properties of the proposed QCP in BaFe$_2$(As$_{1-x}$P$_x$)$_2$ and
its relationship with the observed peak of $\lambda_{L}$.

\begin{figure}
\centering \epsfig{file=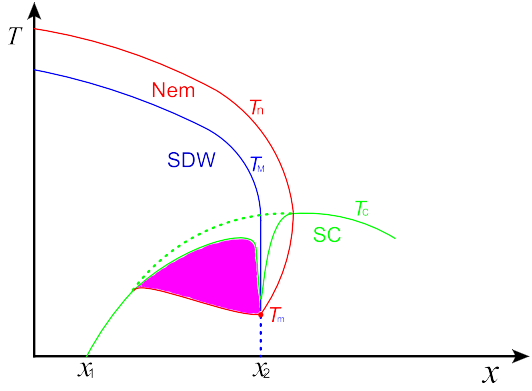,height=5.1cm,width=7cm}
\vspace{-0.35cm} \caption{Schematic phase diagram of iron pnictides
on the $x-T$ plane \cite{Moon2010PRB, Fernandes2013PRL,
Fernandes2014NPhys}, where $x$ denotes the doping concentration. The
two points $x_1$ and $x_2$ are the SC and SDW QCPs, respectively.
The nematic transition line bends back towards lower $x$ in the SC
dome, showing reentrant behavior \cite{Nandi}.}\label{Fig_coexist}
\end{figure}

However, the above finding seems to be at odds with the fact that
there are two transition lines going across the superconducting line
$T_c$, namely a nematic transition line $T_{n}$ and a spin density
wave (SDW) transition line $T_{M}$. This important issue was
addressed by Fernandes \emph{et al.} \cite{Fernandes2013PRL} within
an effective theory that consists of SC, SDW, and nematic order
parameters. The theoretical analysis of Ref.~\cite{Fernandes2013PRL}
demonstrated that the SDW and nematic transition lines penetrate
separately into the SC dome, but merge at certain temperature,
giving rise to a single QCP, which is schematically shown in
Fig.~\ref{Fig_coexist}.

In spite of the interesting progress, our understanding of the
physical effects of ordering competition is still quite limited, and
more research effort is required. In particular, it is necessary to
investigate how the global phase diagram is influenced by ordering
competition, which can help us to clarify many important issues
about the nature of quantum phase transitions in iron pnictides.

In this paper, we study the global phase diagram of some iron
pnictides, such as BaFe$_2$(As$_{1-x}$P$_x$)$_2$ and
Ba(Fe$_{1-x}$Co$_x$)$_2$As$_2$, by carefully investigating the
impact of the competition among superconductivity, SDW order and
nematic order. In order to examine the role played by the quantum
fluctuation of various order parameters, we will perform an
extensive renormalization group (RG) analysis \cite{Wilson1975RMP,
Shankar1994RMP} and obtain the RG equations for all the physical
parameters that are introduced to describe the system. We also
extract the $T$-dependence of the superfluid density $\rho_s(T)$
from the solutions of RG equations, and find that the
superconductivity may be drastically suppressed by the ordering
competition. In addition, the RG results clearly show that the phase
transitions become first order.

Moreover, we have paid special attention to the fate of the
transition line of nematic order in the SC dome. Interestingly, our
RG analysis have discovered that the nematic order can only exist in
an intermediate temperature region $T_{m} < T < T_{n}$, but is
destroyed at $T > T_{n}$ by thermal fluctuation and at $T < T_{m}$
by ordering competition. Such a phenomenon is found to occur in a
wide region of the SC. This result indicates that the nematic
transition line bends back towards lower values of $x$, which is
shown in the schematic phase diagram Fig.~\ref{Fig_coexist}. We
notice that Nandi \emph{et al.} \cite{Nandi} have observed a
reentrance of the tetragonal structure at low $T$ in the SC dome.
This observation is phenomenologically analogous to our theoretical
result about the fate of nematic order.

It is known \cite{Levchenko2013PRL, Fernandes2013PRL} that many of
the physical properties of BaFe$_2$(As$_{1-x}$P$_x$)$_2$ and
$\mathrm{Ba(Fe_{1-x} Co_x)_2 As_2}$ can be described by a three-band
model that composed of one hole pocket located at the center of the
Brillouin zone $\mathbf{Q}_\Gamma = (0,0)$ and two electron pockets
centered at two specific momenta $\mathbf{Q}_X=(\pi,0)$ and
$\mathbf{Q}_Y = (0,\pi)$. The Hamiltonian is usually written as
\cite{Fernandes2013PRL}
\begin{eqnarray}
H=\sum_{\mathbf{k},i\in(X,Y,\Gamma)}\varepsilon_{\mathbf{k},i}
c^\dagger_{\mathbf{k}\sigma,i}c_{\mathbf{k}\sigma,i} + H_4,
\end{eqnarray}
where $H_4$ represents two sorts of interactions:
\begin{eqnarray}
H_4 &=& \sum_{\mathbf{k},i\in(X,Y)}\frac{U_3}{2}
\left(c^\dagger_{\mathbf{k}\alpha,\Gamma}
c^\dagger_{\mathbf{k}\gamma,\Gamma}
c_{\mathbf{k}\delta,i}c_{\mathbf{k}\beta,i} + \mathrm{h.c.}\right)
\delta_{\alpha\beta}\delta_{\gamma\delta}\nonumber\\
&& + \sum_{\mathbf{k},i\in(X,Y)}U_1
c^\dagger_{\mathbf{k}\alpha,\Gamma} c^\dagger_{\mathbf{k}\gamma,i}
c_{\mathbf{k}\delta,i}c_{\mathbf{k}\beta,\Gamma}
\delta_{\alpha\beta}\delta_{\gamma\delta}.
\end{eqnarray}
Here, $U_3$ denotes the pair hoping interaction and $U_1$ the
density-density interaction, which are responsible for the SDW order
and superconductivity respectively \cite{Fernandes2013PRL,
Maiti2010PRB}. We mention here that the $U_3$ terms describe the
magnetic Hund's coupling interactions \cite{Chubukov2008PRB,
Lee2009PRL}, which are known to be important in multi-band
electronic systems \cite{Medici2011PRB, Schickling2012PRL,
Medici2014PRL}. As demonstrated in Ref.~\cite{Fernandes2013PRL}, the
fermionic degrees of freedom can be fully integrated out, leading to
\begin{eqnarray}
\mathcal{L} &=& \frac{1}{2}(\partial_\mu M_X)^2 +
\frac{1}{2}(\partial_\mu M_Y)^2 + a_m(M^2_X + M^2_Y) \nonumber\\
&& + \frac{u_m}{2}(M^2_X + M^2_Y)^2 - \frac{g_m}{2}(M^2_X - M^2_Y)^2
\nonumber\\
&& + \partial_\mu \Delta^\dagger\partial_\mu \Delta +
a_s\Delta^2 + \frac{u_s}{2}\Delta^4
\nonumber\\
&& + \zeta(M^2_X + M^2_Y)\Delta^2,
\end{eqnarray}
where the parameters $a_m$, $a_s$, $u_s$, $u_m$, $g_m$, and $\zeta$
are defined in \cite{Fernandes2013PRL}. This model will be our
starting point. The transition lines for the SDW and SC orders are
obtained by taking $a_m=0$ and $a_s=0$, respectively. $M_{X,Y}$
represent the SDW order parameters that generate long-range magnetic
order for $(\pi,0)$ and $(0,\pi)$ respectively
\cite{Fernandes2012PRB, Fernandes2013PRL}. For $s^{+-}$-wave
superconductor, a universal SC gap is introduced such that
$\Delta_\Gamma = -\sqrt{2}\Delta_{X,Y} = \Delta$
\cite{Fernandes2013PRL}. An Ising-type nematic order is induced by
the magnetic order \cite{Fernandes2012PRB}, and represented by the
$M^2_X - M^2_Y$.

Fernandes \emph{et al.} \cite{Fernandes2013PRL} have recently
studied the nature of quantum phase transitions in iron pnictides
within the above effective theory and argued that the SDW and
nematic orders merge at certain point in the SC dome. In this paper,
we will make a systematic RG analysis. Our aim is two fold. First,
we would like to examine the impact of the quantum fluctuations of
SC and magnetic order parameters on the fate of phase transitions
and the global phase diagram of the system, since these quantum
fluctuations are known to play a vital role in systems that exhibit
competing orders \cite{She2010PRB,Millis2010,Wang2014PRD}. Second, we
attempt to extract the $T$-dependence of the superfluid density $\rho_s(T)$
from the RG solutions. As aforementioned, $\rho_s(T)$ can be
suppressed by two different scenarios: the competitive interaction
between distinct orders, and the coupling between fermionic
quasiparticles and competing order. While the latter scenario has
been studied in some recent references \cite{Levchenko2013PRL,
Ikeda2013PRL}, the former is rarely considered in the literature
\cite{Liu2012PRB}. As will be shown below, the influence of ordering
competition on $\rho_s(T)$ is prominent and can be efficiently
obtained from our RG results.

To simplify consideration, we first concentrate on the SDW QCP,
corresponding to $x_2$ on Fig.~\ref{Fig_coexist}, at which $a_m = 0$
and the magnetic order parameters have vanishing mean values, i.e.,
$\langle M_X \rangle = \langle M_Y \rangle = 0$. In the SC dome, the
SC order parameter develops a nonzero mean value, i.e.,
$\langle\Delta\rangle = V_0 = \sqrt{-a_s/u_s}$. To study the quantum
fluctuation of $\Delta$ around $\langle\Delta\rangle$, we introduce
two new fields $h$ and $\eta$, and then decompose $\Delta$ as
$\Delta = V_0+\frac{1}{\sqrt{2}}(h+i\eta)$ with $\langle h
\rangle=\langle \eta \rangle=0$. Moreover, to compute the superfluid
density, it is convenient to couple $\Delta$ to a gauge potential
$A$. Recall that an important property of SC state is the happening
of Anderson-Higgs mechanism, which leads to the Meissner effect.
This mechanism needs to be properly included in the RG analysis. We
now introduce gauge potential $A$ to the effective theory via the
standard minimal coupling, i.e., $\partial_{\mu}\Delta\rightarrow
(\partial_{\mu}-i\zeta_{\Delta A}A_{\mu})\Delta$, where
$\zeta_{\Delta A}$ is the coupling between $A$ and $\Delta$.
Straightforward calculations gives rise to the following effective
model
\begin{eqnarray}
\mathcal{L}_{\mathrm{eff}} &=& \frac{1}{2}(\partial_\mu M_X)^2 +
\frac{1}{2}(\partial_\mu M_Y)^2 + \alpha_X M^2_X + \alpha_Y M^2_Y
\nonumber \\
&& + \frac{\beta_X}{2}M^4_X + \frac{\beta_Y}{2}M^4_Y +
\zeta_{XY}M^2_X M^2_Y \nonumber \\
&& + \frac{1}{2}(\partial_\mu h)^2 + \alpha_hh^2 +
\frac{\beta_h}{2}h^4 + \gamma_h h^3 \nonumber \\
&& -\frac{1}{4}\left(\partial_\mu A_\nu - \partial_\nu
A_\mu\right)^2 + \frac{\alpha_A}{2}A^2 \nonumber \\
&& + \gamma_{X^2h}M^2_Xh + \gamma_{Y^2h}M^2_Yh +
\zeta_{Xh}M^2_Xh^2 \nonumber \\
&& + \zeta_{Yh}M^2_Yh^2 + \gamma_{hA^2}hA^2 +
\zeta_{hA}h^2A^2,\label{Eq_effective_L}
\end{eqnarray}
where we have defined a number of new effective parameters from
$a_m$, $a_s$, $u_s$, $u_m$, $g_m$, $\zeta$, and $\zeta_{\Delta A}$:
\begin{eqnarray}
&&\alpha_X = \alpha_Y \equiv a_m - \frac{a_s\zeta}{u_s}, \beta_X =
\beta_Y \equiv u_m - g_m,\nonumber \\
&&\alpha_h \equiv - a_s, \beta_h \equiv \frac{u_s}{4}, \gamma_h
\equiv \sqrt{\frac{-a_su_s}{2}}, \alpha_A \equiv -
\frac{2a_s\zeta_{\Delta A}}{u_s},\nonumber \\
&&\gamma_{hA^2}\equiv \zeta_{\Delta A}\sqrt{\frac{-2a_s}{u_s}},
\gamma_{X^2h} = \gamma_{Y^2h} \equiv
\zeta\sqrt{\frac{-2a_s}{u_s}}, \nonumber \\
&& \zeta_{XY} \equiv (u_m+g_m), \zeta_{Xh} = \zeta_{Yh} \equiv
\frac{\zeta}{2}, \zeta_{hA} \equiv \frac{\zeta_{\Delta A}}{2},
\label{Eq_para_trans}
\end{eqnarray}
As a consequence of the Anderson-Higgs mechanism, the massless
Goldstone boson induced by gauge symmetry breaking is swallowed by
the gauge boson $A$, which then acquires an effective mass term
$\frac{1}{2}\alpha_A A^2$. The superfluid density $\rho_s$ is
proportional to the gauge boson mass, i.e., $\rho_s \propto
\alpha_A$ \cite{Halperin1974PRL}. In the following, we will extract
the $T$-dependence of $\rho_s(T)$ by computing the $l$-dependence of
$\alpha_A \equiv \alpha_A(l)$, where $l$ is a running length scale.
However, $\alpha_A$ is related intimately to other parameters, so we
need to solve all the RG equations self-consistently.

\begin{figure}
\centering \epsfig{file=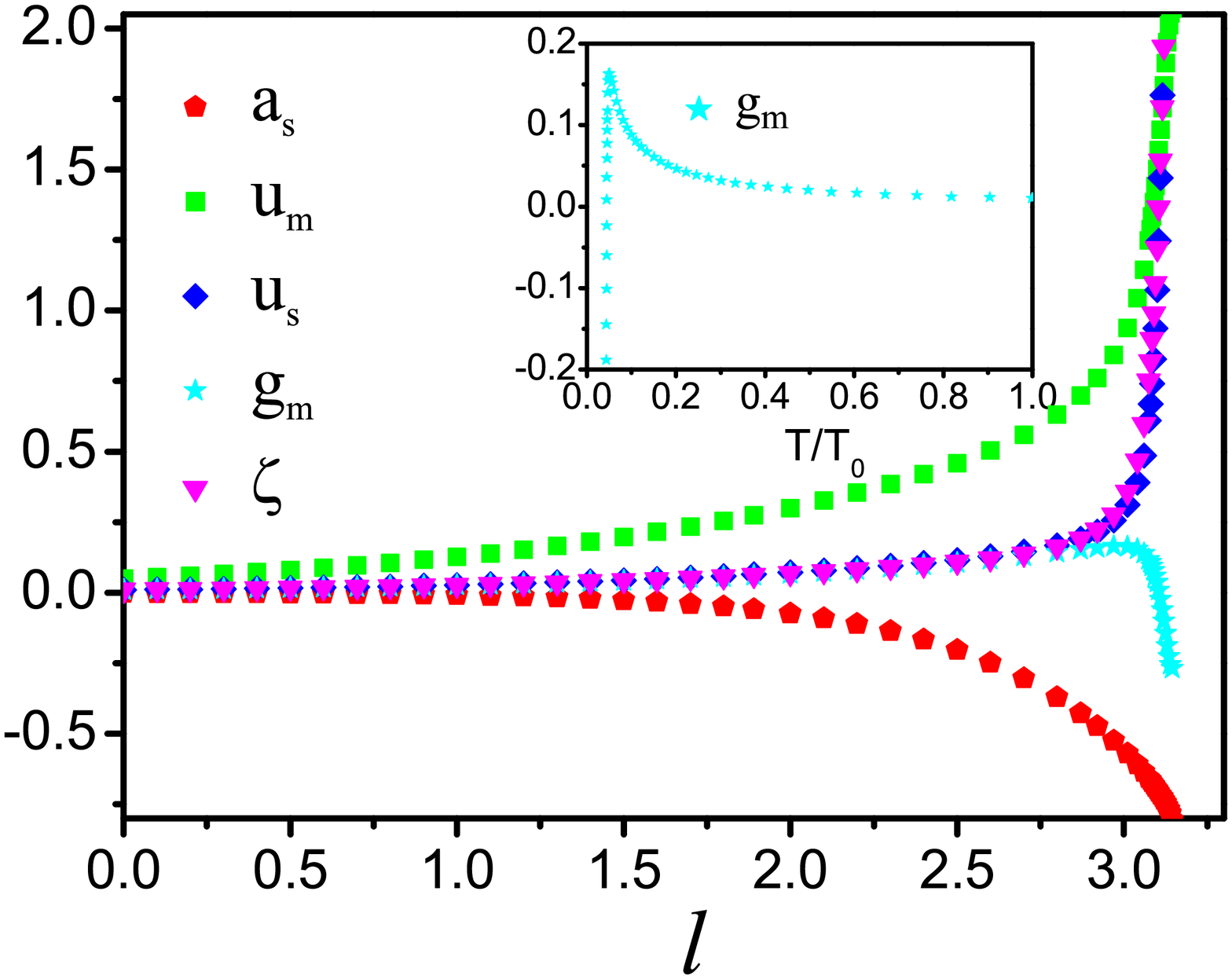,height=6.2cm,width=9.3cm}
\vspace{-0.69cm} \caption{Flows of $a_s$, $u_m$, $u_s$, $g_m$ and
$\zeta$ and $T$-dependence of $g_m$ (inset) at SDW QCP. The
parameter $g_m$ is positive at small $l$ and high $T$, but becomes
negative suddenly once $l$ exceeds some critical value $l_m$ and $T$
is lower than $T_m$.}\label{Fig_runaway_gm_T}
\end{figure}

As shown in Eq.~(\ref{Eq_para_trans}), all the new effective
parameters are given by the seven original parameters. Our RG
calculations are performed to the one-loop level in powers of the
coupling constants. Analogous to the analysis of
Ref.~\cite{Wang2014PRD}, we have derived the RG equations for all
the coupling parameters (See Supplemental Material \cite{Suppl} for
the details).

To solve the equations, we assign the initial values of parameters
as $a_s=-0.001,u_m=0.05, u_s=0.01,g_m=0.01,\zeta = 0.01$, and
$\zeta_{\Delta A} = 1.0\times10^{-8}$. SC and SDW orders are
supposed to coexist, so $u_s$, $u_m$, $g_m$, and $\zeta$ satisfy the
constraint $\zeta < \sqrt{u_s(u_m-g_m)}$ \cite{Schmalian2010PRB,
Fernandes2013PRL}. The main conclusion is independent of these
assumptions. There are three main results, to be explained one by
one below.

First of all, we consider the fate of the associated phase
transitions. After solving the RG equations, we show the
$l$-dependence of various parameters in Fig.~\ref{Fig_runaway_gm_T}.
The quadratic coupling parameters $u_m$, $u_s$, and $\zeta$ all flow
to infinity eventually as $l \rightarrow +\infty$, thus the system
does not approach any stable fixed point in the low energy region.
As discussed in previous works \cite{She2010PRB, Millis2010,
Wang2014PRD, Domany_Chen_Rudnick_Iacobson}, this results is usually
regarded as a signature of an instability towards first-order
transitions since a second-order transition is always associated
with the presence of a stable infrared fixed point
\cite{Wilson1975RMP}. However, we can see from
Fig.~\ref{Fig_runaway_gm_T} that the parameters $u_m$, $u_s$, and
$\zeta$ increase with growing $l$ very slowly for small values of
$l$. Their magnitudes become large only when $l$ grows beyond
certain threshold. Therefore, before running to large values, the RG
results of the $l$-dependence of these parameters are still reliable
and give us very useful information about the physical properties of
the system \cite{Wang2014PRB}.

\begin{figure}
\centering \epsfig{file=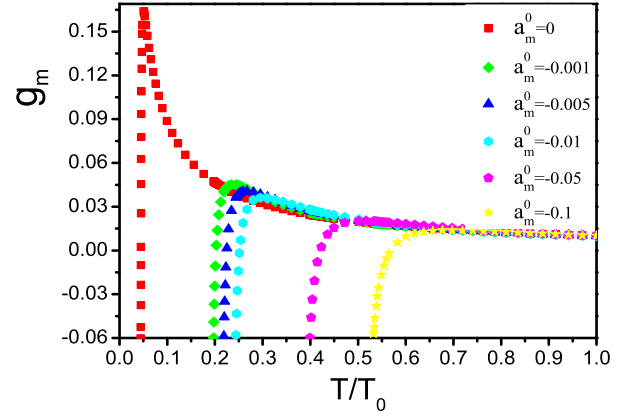,height=6.2cm,width=9.3cm}
\vspace{-0.69cm} \caption{$T$-dependence of $g_m$ for different
values of bare parameter $a^0_m$, which measures the distance to SDW
QCP $x_2$. As $a_m^0$ grows, $g_m$ varies with $T$ similarly, but
its peak disappears and the temperature for $g_m$ to change sign
increases.} \label{Fig_g_m_II}
\end{figure}

Secondly, we address the impact of the ordering competition on the
nematic transition line. A number of experiments have observed an
interesting transition from an orthorhombic structure to a
tetragonal structure at low $T$ in the SC dome of
$\mathrm{Ba(Fe_{1-x} Co_x)_2 As_2}$ \cite{Nandi, Pratt2009PRL,
Christianson2009PRL} and $\mathrm{Ba(Fe_{1-x} Rh_x)_2 As_2}$
\cite{Kreyssig2010PRB}. It turns out that the nematic order exists
in an intermediate range of $T$ and disappears once $T$ is lower
than certain value. Remarkably, this unusual behavior can be
naturally obtained in our RG analysis. To demonstrate this, let us
consider the property of parameter $g_m$, whose sign determines
whether the nematic order is present. If $g_m
> 0$, only one of the two order parameters $M_X$ and $M_Y$ develops
a finite mean value \cite{Fernandes2012PRB, Fernandes2014NPhys,
Fernandes1504.03656} due to tetragonal symmetry breaking. In this
case, the nematic order is present. On the other hand, we have
$\langle M_X \rangle = \langle M_Y \rangle$ if $g_m < 0$, which
indicates that the nematic order is absent \cite{Fernandes2012PRB,
Fernandes2014NPhys, Fernandes1504.03656}. Therefore, to judge
whether the nematic order exists at certain $T$, we need to compute
the $T$-dependence of $g_m$ from the RG results.

The detailed $l$-dependence of $g_m$ is depicted in
Fig.~\ref{Fig_runaway_gm_T}. As $l$ grows, $g_m$ first increases
steadily, but decreases rapidly for large values of $l$ and becomes
negative at certain critical value $l_m$. Notice that $l_m$ is also
the length scale at which $u_m$, $u_s$, and $\zeta$ diverge. We can
translate the $l$-dependence of $g_m$ to a $T$-dependence by using
the transformation \cite{She2015PRB} $T = T_0 e^{-l}$. The inset of
Fig.~\ref{Fig_runaway_gm_T} clearly shows that the positive $g_m$
becomes negative as $T$ decreases immediately below temperature
$T_{m} = T_0 e^{-l_m}$, which means the nematic order is entirely
suppressed at $T < T_{m}$. Therefore, the nematic order can only
exist in the intermediate range between $T_m$ and its transition
temperature $T_n$. It is destroyed by the thermal fluctuation at $T
> T_n$, and by the strong competition among superconductivity, SDW,
and nematic order at $T < T_m$. Since $T_n$ is supposed to be always
higher than $T_{M}$ \cite{Fernandes2013PRL}, this behavior gives
rise to a back-bending effect of the nematic transition line on the
phase diagram shown in Fig.~\ref{Fig_coexist}, which is consistent
with the observed reentrance of tetragonal structure at low
temperatures \cite{Nandi}.

\begin{figure}
\centering \epsfig{file=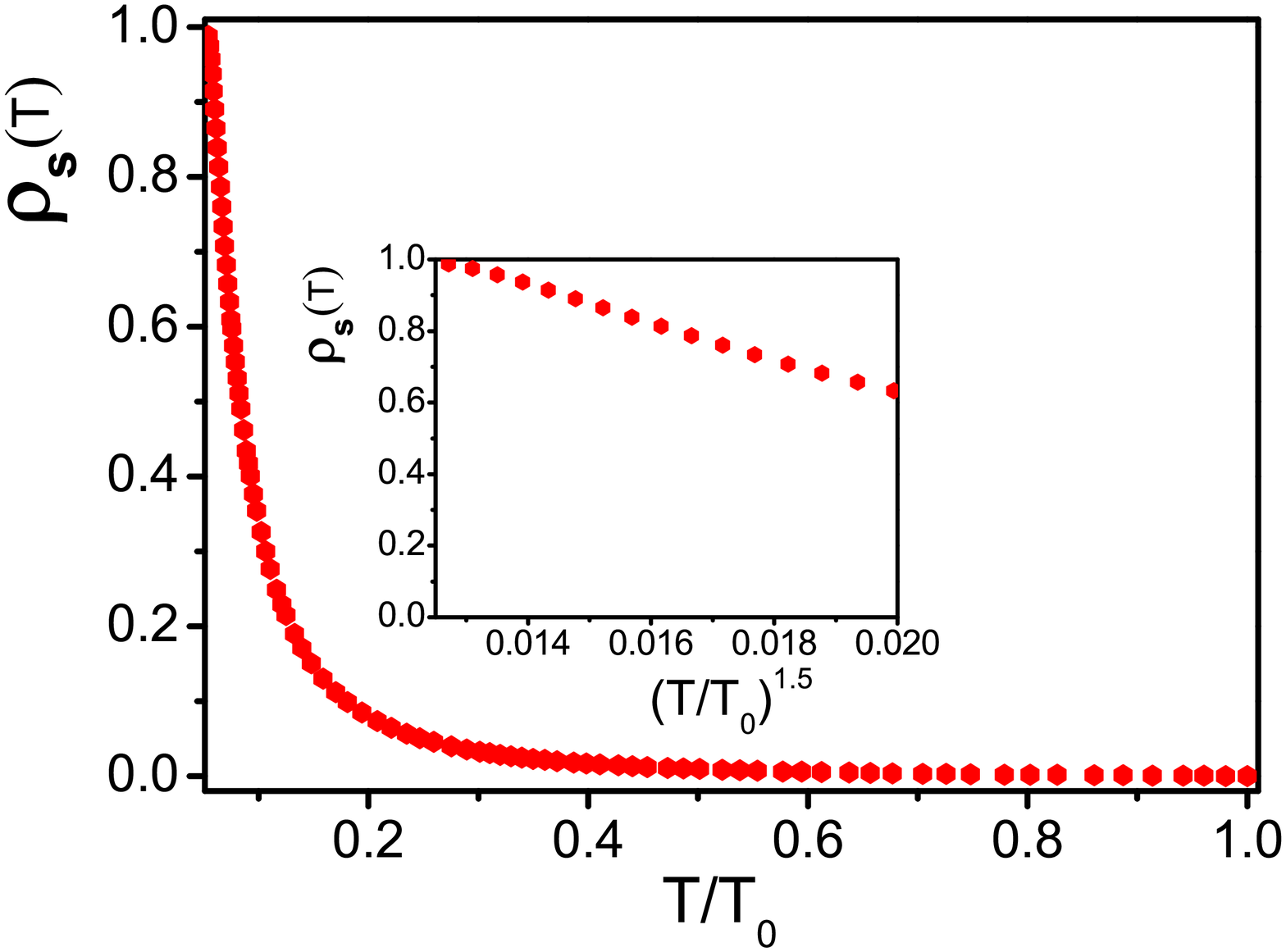,height=6.2cm,width=9.3cm}
\vspace{-0.69cm} \caption{Strong $T$-dependence of superfluid
density $\rho_s(T)$ at the SDW QCP. The inset shows that $\rho_s(T)$
is similar to $(T/T_0)^{1.5}$ only in a very restricted temperature
range.}\label{Fig_denisty_L}
\end{figure}

To acquire a better knowledge of the phase diagram, we also wish to
know how the nematic transition line varies with $T$ as we move away
from the SDW QCP. In the doping region $x_1 < x < x_2$, the SDW
order parameters $M_X$ and $M_Y$ develop finite mean values. We here
only present the main results (See Supplemental Material
\cite{Suppl} for more details). Fig.~\ref{Fig_g_m_II} shows the
$T$-dependence of $g_m$ for different values of bare parameter
$a_m^0$. For the chosen values of $a_m^0$, $g_m$ always first grows
with decreasing $T$ and then becomes negative once $T$ is below
certain threshold, which means the nematic order is suppressed at
low $T$. In addition, the temperature scale at which $g_m$ changes
sign increases as $a_m^0$ grows. In principle, the quantity $T_0$,
given by the SC transition temperature, also varies as $a_m^0$
grows. However, despite this complexity, one can conclude from our
RG results that the nematic transition line cannot intersect with
the horizontal axis of Fig.~\ref{Fig_coexist}, but should instead
merge somewhere with the SC transition line. This property leads to
a considerable modification of the global phase diagram, as
visualized in Fig.~\ref{Fig_coexist}. According to our results, the
reentrance of tetragonal phase occur in a wide range of doping $x$.

\begin{figure}
\centering \epsfig{file=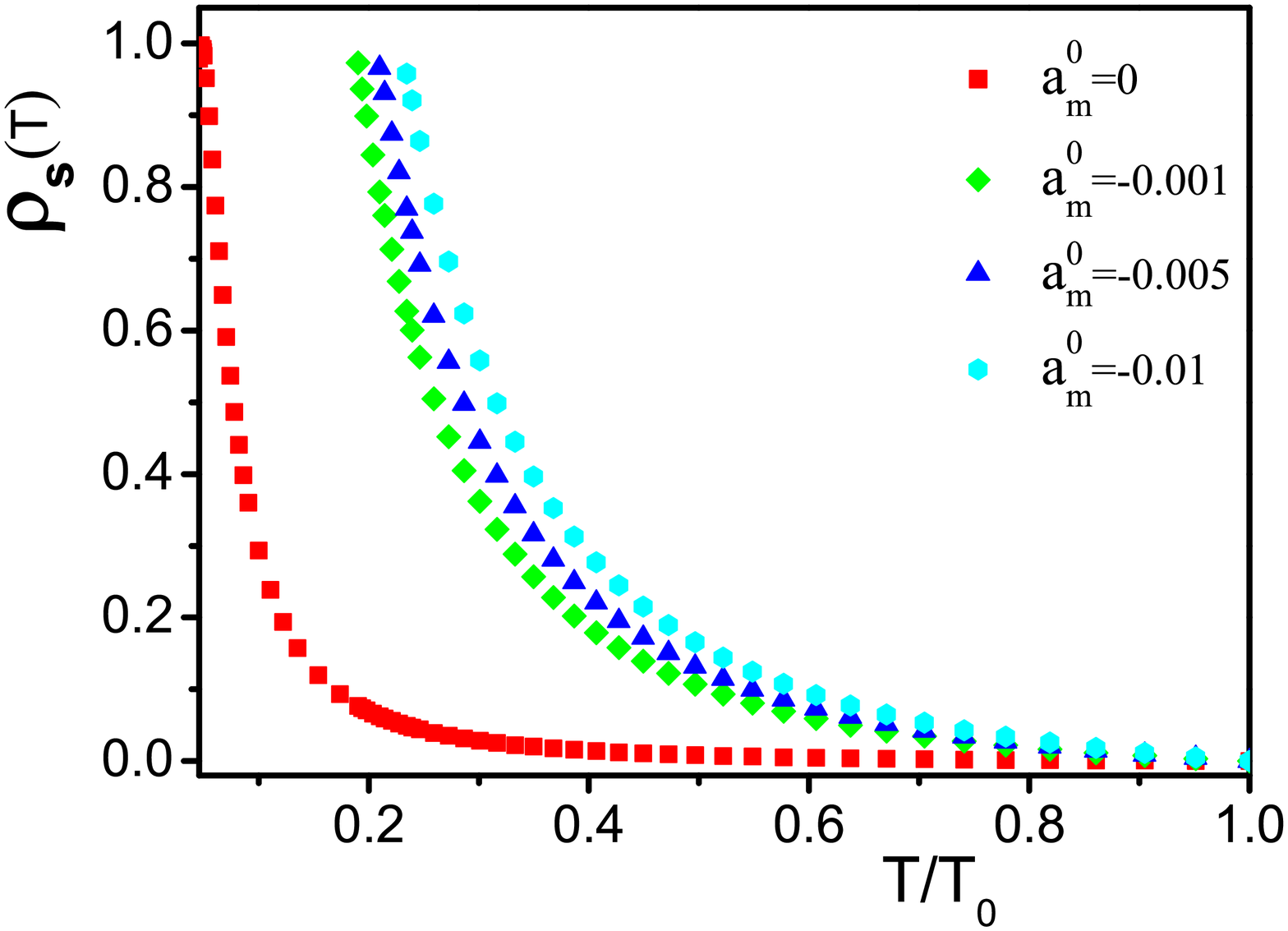,height=6.2cm,width=9.3cm}
\vspace{-0.69cm} \caption{$\rho_s(T)$ at different values of
$a_m^0$. The suppression of superfluid density takes place for any
$a_m^0$, but is most significant at the SDW QCP where $a_m^0 = 0$.}
\label{Fig_density_II}
\end{figure}

Finally, we turn to the $T$-dependence of superfluid density
$\rho_s$. Within the effective model given by
Eq.~(\ref{Eq_effective_L}), the superfluid density $\rho_s \propto
\alpha_A=-\frac{2a_s\zeta_{\Delta A}}{u_s}$. To obtain the
$T$-dependent $\lambda_{L}(T)$, we also utilize the transformation
$T = T_0e^{-l}$, where a suitable choice of $T_0$ is the SC
temperature $T_{c}$. At first glance, the RG results seem to suggest
that $\rho_s \propto \lambda_L^{-2}$ diverges rapidly at $T_m$.
However, because the transitions become first order, the
$T$-dependence of $\rho_s(T)$ is reliable only at $T > T_m$. Here,
we choose the value $\rho_s(T^{\ast})$ with $T^{\ast}$ being a
little higher than $T_m$ to re-scale $\rho_s(T)$, and then show
$\rho_s(T)/\rho(T^{\ast})$ in Fig.~\ref{Fig_denisty_L}, where the
initial values are the same as those adopted in
Fig.~\ref{Fig_runaway_gm_T}. An obvious conclusion is that the
ordering competition leads to a strong $T$-dependence of $\rho_s$,
which decreases very rapidly as $T$ grows. For finite $a_m^0$, the
behaviors of $\rho_s(T)$ are depicted in Fig.~\ref{Fig_density_II}.
As $a_m^0$ increases, the suppression of $\rho_s(T)$ becomes weaker.
Since $T_c$ sensitively depends on the superfluid density, we can
infer that $T_c$ should be suppressed to some extent in the region
$x_1 < x < x_2$ and that this effect is most significant at $x_2$,
as shown in Fig.~\ref{Fig_coexist}.

Hashimoto \emph{et al.} \cite{Hashimoto2012PNAS} has measured the
superfluid density $\rho_s(T)$ in a number of superconductors,
including iron pnictides and heavy fermion compounds, and claimed to
unveil a universal $\rho_s(T) \propto T^{1.5}$ behavior over a wide
range of temperatures. This behavior was argued
\cite{Hashimoto2012PNAS, Ikeda2013PRL} to be caused by the coupling
between magnetic fluctuation and fermionic excitations. Our analysis
has clearly showed that ordering competition alone cannot account
for the $\rho_s(T) \propto T^{1.5}$ behavior. As depicted in the
inset of Fig.~\ref{Fig_denisty_L}, only in a very restricted
region of $T$ could we extract an approximate $(T/T_0)^{1.5}$
behavior. On the other hand, we can see from the main panel of
Fig.~\ref{Fig_denisty_L} that, ordering competition does make a
significant contribution to $\rho_s(T)$ and hence cannot be simply
neglected. Bearing these two points in mind, we believe that both
ordering competition and fermionic excitations need to be properly
incorporated in a more refined model of $\rho_s(T)$.

In summary, we have studied the impact of the strong competition
among superconductivity, SDW order, and nematic order on the global
phase diagram of iron pnictides by performing a systematic RG
analysis within an effective field theory. The main results are
summarized in the schematic phase diagram presented in
Fig.~\ref{Fig_coexist}.

J.W. is supported by the China Postdoctoral Science Foundation under
Grants 2015T80655 and 2014M560510, the National Natural Science Foundation
of China under Grant 11504360 and the Fundamental Research Funds for the
Central Universities (P. R. China) under Grant WK2030040074. G.Z.L. is
supported by the National Natural Science Foundation of China under
Grant 11274286.

%%%%%%%%%%%%%%%%%%%%%%%%%%%%%%%%%%%%%%%%%%%%%%%%%%%%%%%%%%%%%%%%%%%%%%%%%%%%%

\newpage

\begin{widetext}

\section{Supplementary Material: Impact of ordering competition on
the global phase diagram of iron pnictides}

The 122-family iron-based superconductors, such as
BaFe$_2$(As$_{1-x}$P$_x$)$_2$ and $\mathrm{Ba(Fe_{1-x} Co_x)_2
As_2}$ are widely described in the literature
\cite{Levchenko2013PRL, Fernandes2013PRL} by a three-band model that
is composed of one hole pocket located at the center of the
Brillouin zone $\mathbf{Q}_\Gamma = (0,0)$ and two electron pockets
centered at two specific momenta $\mathbf{Q}_X=(\pi,0)$ and
$\mathbf{Q}_Y = (0,\pi)$. Recently, this model was used by Fernandes
\emph{et al.} to study the nature of quantum phase transitions in
the superconducting dome of some iron pnictides
\cite{Fernandes2013PRL}. After integrating out the fermionic degrees
of freedom \cite{Fernandes2013PRL} and including the kinetic terms,
one can obtain an effective Lagrangian density
\begin{eqnarray}
\mathcal{L} &=& \frac{1}{2}(\partial_\mu M_X)^2 +
\frac{1}{2}(\partial_\mu M_Y)^2 + a_m(M^2_X + M^2_Y) +
\frac{u_m}{2}(M^2_X + M^2_Y)^2 - \frac{g_m}{2}(M^2_X - M^2_Y)^2
\nonumber\\
&& + \partial_\mu \Delta^\dagger\partial_\mu \Delta + a_s\Delta^2 +
\frac{u_s}{2}\Delta^4 + \zeta(M^2_X + M^2_Y)\Delta^2,\label{Eq_L}
\end{eqnarray}
where the parameters $a_m$, $a_s$, $u_s$, $u_m$, $g_m$, and $\zeta$,
are defined in Ref.~\cite{Fernandes2013PRL}. Our RG analysis starts
from this effective field theory. In order to examine the effect of
ordering competition on the superfluid density, we introduce a small
external gauge potential $A$ that couples to the superconducting
order parameter \cite{Halperin1974PRL} and thus have an additional
term
\begin{eqnarray}
\mathcal{L'} &=& -\frac{1}{4}\left(\partial_\mu A_\nu-\partial_\nu
A_\mu\right)^2 + \zeta_{\Delta A}|\Delta|^2 A^2,\label{Eq_L2}
\end{eqnarray}
where the Lorentz gauge $\partial_{\mu}A_{\mu} = 0$ is utilized and
the parameter $\zeta_{\Delta A}$ is a new coupling constant.

In the superconducting dome, the order parameter $\Delta$ acquires a
finite vacuum expectation value, i.e.,
\begin{eqnarray}
V_0\equiv\langle \Delta\rangle=\sqrt{\frac{-a_s}{u_s}}.
\end{eqnarray}
The quantum fluctuation of $\Delta$ around its mean value is
believed to play an important role \cite{Wang2014PRD} and needs to
be seriously taken into account. It is convenient to introduce two
new fields $h$ and $\eta$ by defining
\begin{eqnarray}
\Delta = V_0 + \frac{1}{\sqrt{2}}(h +
i\eta),\label{Eq_parametrization}
\end{eqnarray}
with $\langle h \rangle = \langle \eta \rangle = 0$. In many
field-theoretic treatments of ordering competition, specially in the
context of condensed matter systems, the quantum fluctuation of
order parameter in the ordered phase is usually omitted. However,
more careful analysis \cite{Wang2014PRD} have showed that this
approximation is not appropriate and that the order parameter
fluctuation around its mean value can be significant. In order to
entirely reveal the physical effects of ordering competition, we
will consider $h$ and $\eta$ as quantum field operators and study
their interactions with the magnetic order parameters $M_X$ and
$M_Y$.

Substituting the re-parameterized field operator
Eq.~(\ref{Eq_parametrization}) into the total Lagrangian density, we
get a new effective model
\begin{eqnarray}
\mathcal{L}_{\mathrm{eff}} &=& \frac{1}{2}(\partial_\mu
M_X)^2+\frac{1}{2}(\partial_\mu M_Y)^2 + \alpha_X M^2_X + \alpha_Y
M^2_Y + \frac{\beta_X}{2}M^4_X +
\frac{\beta_Y}{2}M^4_Y + \zeta_{XY}M^2_X M^2_Y\nonumber\\
&&+\frac{1}{2}(\partial_\mu h)^2 + \alpha_h h^2 +
\frac{\beta_h}{2}h^4 + \gamma_h h^3 - \frac{1}{4}\left(\partial_\mu
A_\nu - \partial_\nu A_\mu\right)^2 + \frac{\alpha_A}{2}A^2 +
\gamma_{X^2h}M^2_X h \nonumber \\
&&+\gamma_{Y^2h}M^2_Y h + \zeta_{Xh}M^2_X h^2 + \zeta_{Yh}M^2_Y h^2
+ \gamma_{h A^2}h A^2 + \zeta_{hA}h^2 A^2,\label{Eq_L_eff}
\end{eqnarray}
where a number of effective parameters are defined on the basis of
the fundamental parameters $a_m$, $a_s$, $u_s$, $u_m$, $g_m$,
$\zeta$, and $\zeta_{\Delta A}$, as given by Eq.~(5) in the main
part of the paper. Our RG analysis will be based on this effective
model, assuming that the coupling constants take small values.

We now proceed to treat the effective theory by performing a
detailed RG analysis \cite{Shankar1994RMP}. To this end, we first
make the following scaling transformations
\begin{eqnarray}
k_i &=& k'_ie^{-l},\nonumber\\
\omega &=& \omega' e^{-l},\nonumber\\
q_i &=& q'_i e^{-l},\nonumber\\
\epsilon &=& \epsilon' e^{-l},\label{Eq_scaling}
\end{eqnarray}
where $i = x,y$ and $l$ is a running length scale. Under these
transformations, the field operators $M_X$, $M_Y$, $h$, and $\eta$
should be re-scaled as
\begin{eqnarray}
M_X(\mathbf{k},\omega) &=& M'_X(\mathbf{k'},\omega')e^{5l/2},\nonumber\\
M_Y(\mathbf{k},\omega) &=& M'_Y(\mathbf{k'},\omega')e^{5l/2},\nonumber\\
h(\mathbf{k},\omega) &=& h'(\mathbf{k'},\omega')e^{5l/2},\nonumber\\
A(\mathbf{q},\epsilon) &=& A'(\mathbf{q'},\epsilon')e^{5l/2}.
\end{eqnarray}

In order to obtain the flow equations of the fundamental parameters
defined in Eq.~(\ref{Eq_L}) and Eq.~(\ref{Eq_L2}), we apply the
following identifies:
\begin{eqnarray}
\frac{da_s}{dl} &=& -\frac{d\alpha_h}{dl}, \nonumber \\
\frac{du_s}{dl} &=& 4\frac{d\beta_h}{dl}, \nonumber \\
\frac{da_m}{dl} &=& \frac{d\alpha_X}{dl} + \frac{\zeta}{u_s}
\frac{da_s}{dl} + \frac{a_s}{u_s}\frac{d\zeta}{dl} -
\frac{a_s\zeta}{u_s^2}\frac{du_s}{dl},\nonumber \\
\frac{du_m}{dl} &=& \frac{1}{2}\frac{d\zeta_{XY}}{dl} +
\frac{1}{2}\frac{d\beta_X}{dl},\nonumber \\
\frac{dg_m}{dl} &=& \frac{1}{2}\frac{d\zeta_{XY}}{dl} -
\frac{1}{2}\frac{d\beta_X}{dl},\nonumber \\
\frac{d\zeta}{dl} &=& 2\frac{d\zeta_{Xh}}{dl}, \nonumber
\\
\frac{d\zeta_{\Delta A}}{dl} &=& 2\frac{d\zeta_{hA}}{dl}.
\end{eqnarray}

For simplicity, we first consider the SDW QCP with $a_m = 0$, where
the magnetic order parameters $M_X$ and $M_Y$ both have vanishing
mean values and their quantum fluctuations are critical. Analogous
to the scheme presented in Ref.~\cite{Wang2014PRD}, we have derived
the following RG equations for the seven fundamental parameters
\begin{eqnarray}
\left.\begin{array}{ll} \frac{da_s}{dl} = 2a_s -
\frac{1}{12\pi^2}\left[\frac{27a_su_s}{2}(1+4a_s)+\frac{12a_s\lambda^2}{u_s}\left(1+
\frac{4a_s\lambda}{u_s}\right)+3\lambda\left(1+\frac{2a_s\lambda}{u_s}\right)\right.\\
\hspace{0.77cm}\left.+\frac{9u_s}{4}(1+2a_s)+3\lambda_{\Delta A}
\left(1+\frac{2a_s\lambda_{\Delta
A}}{u_s}\right)+\frac{32a_s\lambda^2_{\Delta A}}{u_s}
\left(1+\frac{4a_s\lambda_{\Delta A}}{u_s}\right)\right].
\\
\frac{du_m}{dl}
=u_m+\frac{1}{2\pi^2}\left\{(8u_mg_m-17u^2_m-11g^2_m)\left(\frac{4a_s\lambda}
{u_s}+1\right)-\frac{3\lambda^2}{8}(4a_s+1)\right.\\
\hspace{0.77cm}\left.+\frac{16a_s\lambda^2}{u_s}(2u_m-g_m)
\left(1+\frac{4a_s\lambda}{u_s}+2a_s\right)-\frac{6a_s\lambda^3}{u_s}
\left[1+2\left(\frac{a_s\lambda}{u_s}+2a_s\right)\right]\right\},
\\
\frac{du_s}{dl}
=u_s+\frac{1}{\pi^2}\left\{2\lambda^2\left[4\left(a_m-\frac{a_s\lambda}{u_s}\right)-1\right]
-\frac{9u^2_s}{4}(4a_s+1)+54a_su^2_s(1+6a_s)\right.\\
\hspace{0.77cm}\left.-\frac{4\lambda^2_{\Delta A}}{3}
\left(\frac{4a_s\lambda_{\Delta A}}{u_s}+1\right) +
\frac{32a_s\lambda^3}{u_s} \left(1+\frac{6a_s\lambda}{u_s}\right) +
\frac{11072a_s\lambda^3_{\Delta A}}{35u_s}
\left(1+\frac{6a_s\lambda_{\Delta A}}{u_s}\right)\right\},
\\
\frac{dg_m}{dl}
=g_m+\frac{1}{2\pi^2}\left\{3(u_m^2+3g^2_m-8u_mg_m)\left(\frac{4a_s\lambda}{u_s}+1\right)
+\frac{\lambda^2}{8}(4a_s+1)\right.\\
\hspace{0.77cm}\left.+\frac{16a_s\lambda^2}{u_s}(2g_m-u_m)\left(1+\frac{4a_s\lambda}{u_s}
+2a_s\right)-\frac{2a_s\lambda^3}{u_s}\left[1+2\left(\frac{a_s\lambda}
{u_s}+2a_s\right)\right]\right\},
\\
\frac{d\lambda}{dl} = \lambda + \frac{1}{\pi^2}
\left[\frac{4a_s\lambda^3}{u_s}\left(1+\frac{4a_s\lambda}{u_s}+2a_s\right)
-\frac{3u_s\lambda}{8}(4a_s+1)+9a_su_s\lambda(1+6a_s)-2\lambda^2
\left(\frac{2a_s\lambda}{u_s}+2a_s+1\right)\right.\\
\hspace{0.77cm}\left.-\lambda(2u_m-g_m)\left(\frac{4a_s\lambda}{u_s}+1\right)
+\frac{16a_s\lambda^2}{u_s}(2u_m-g_m)\left(1+\frac{6a_s\lambda}{u_s}\right)
+6a_s\lambda^2\left(1+\frac{2a_s\lambda}{u_s}+4\alpha_s\right)\right],
\\
\frac{d\lambda_{\Delta A}}{dl} =\lambda_{\Delta A}+\frac{1}{3\pi^2}
\left\{27a_su_s\lambda_{\Delta A}(1+6a_s)-12\lambda^2_{\Delta A}
\left(1+2a_s+\frac{2a_s\lambda_{\Delta A}}{u_s}\right)-\frac{9u_s\lambda_{\Delta A}}{8}(4a_s+1)\right.\\
\hspace{0.77cm}\left.+\frac{64a_s\lambda^3_{\Delta A}}{u_s}
\left(1+2a_s+\frac{4a_s\lambda_{\Delta
A}}{u_s}\right)+36a_s\lambda^2_{\Delta A}
\left[1+2a_s\left(2+\frac{\lambda_{\Delta
A}}{u_s}\right)\right]\right\}.
\end{array}\right.
\end{eqnarray}

When the system moves away from the SDW QCP and goes to a lower
doping concentration, the stripe-SDW order parameters $M_X$ and
$M_Y$ also develop nonzero mean values, i.e.,
\begin{eqnarray}
\langle M_X\rangle=\langle M_Y\rangle=\sqrt{-\frac{a_m}{2u_m}}.
\end{eqnarray}
To include the quantum fluctuation of $M_X$ and $M_Y$ around their
mean values, we introduce two new fields $\xi$ and $\chi$:
\begin{eqnarray}
M_X &=& \sqrt{-\frac{a_m}{2u_m}}+\xi, \\
M_Y &=& \sqrt{-\frac{a_m}{2u_m}}+\chi,
\end{eqnarray}
with $\langle \xi \rangle = \langle \chi \rangle = 0$. Now the
problem becomes more complicated than the case of SDW QCP. After
lengthy but straightforward calculations, we obtain a set of RG
equations:
\begin{eqnarray}
\left.\begin{array}{ll} \frac{da_s}{dl}
=2a_s-\frac{1}{4\pi^2}\Bigl\{\frac{18a_su_s}{4}(1+4a_s)+\frac{4a_s\lambda^2}{u_s}
\Bigl[1+4\Bigl(\frac{a_s\lambda}{u_s}+a_m\Bigr)\Bigr]+\frac{32a_s\lambda^2_{\Delta
A}}{3u_s} \Bigl(1+\frac{4\lambda_{\Delta A}a_s}{u_s}\Bigr)
+\lambda\Bigl[1+2\Bigl(\frac{a_s\lambda}
{u_s}+a_m\Bigr)\Bigr]\\
\hspace{0.77cm}+\frac{3u_s}{4}(1+2a_s)+\frac{8a_sa^2_m\lambda^2}{u_s}
\Bigl[1+8\Bigl(\frac{a_s\lambda}{u_s}+a_m\Bigr)\Bigr]
+\lambda_{\Delta A}\Bigl(1+\frac{2\lambda_{\Delta A}a_s}{u_s}\Bigr)
+4a^2_m\lambda\Bigl[1+6\Bigl(\frac{a_s\lambda}{u_s}+a_m\Bigr)\Bigr]\Bigr\},\\
\frac{da_m}{dl} = 2\Bigl(\frac{a_s\lambda}{u_s} +
a_m\Bigr)+\frac{1}{4\pi^2}\Bigl\{12a_m\Bigl[a_m(g_m+5u_m)-6u_m\Bigr]
\Bigl[1+4\Bigl(\frac{a_s\lambda}{u_s}+a_m\Bigr)\Bigr]-72a^3_mu_m
\Bigl[1+8\Bigl(\frac{a_s\lambda}{u_s}+a_m\Bigr)\Bigr]\\
\hspace{0.77cm}-\frac{4a_s\lambda^2}{u_s}
\Bigl[1+2\Bigl(\frac{a_s\lambda}{u_s}+a_m\Bigr)+2a_s\Bigr]
-2(2u_m-g_m)\Bigl[1+2\Bigl(\frac{a_s\lambda}{u_s}+a_m\Bigr)\Bigr]-\frac{\lambda}{2}(1+2a_s)\\
\hspace{0.77cm}-4a^2_m(g_m-17u_m)\Bigl[1+6\Bigl(\frac{a_s\lambda}{u_s}
+ a_m\Bigr)\Bigr]\Bigr\} + \left(\frac{\lambda}{u_s}\frac{da_s}{dl}
+ \frac{a_s}{u_s}\frac{d\lambda}{dl} - \frac{a_s\lambda}{u_s^2}\frac{du_s}{dl}\right),\\
\frac{du_m}{dl} = u_m + \frac{1}{4\pi^2} \Bigl\{576a_m u_m(2u_m -
g_m) \left[1+6\Bigl(\frac{a_s\lambda}{u_s}+a_m\Bigr)\Bigr] -
16a^2_m\Bigl(130u^2_m - 94u_m g_m + g^2_m\Bigr)
\Bigl[1+8\Bigl(\frac{a_s\lambda}{u_s}+a_m\Bigr)\Bigr]\right.\\
\hspace{0.77cm}+\frac{32a_s\lambda^2}{u_s}(2u_m-g_m)\Bigl[(1+4a^2_m)(1+2a_s)+4\Bigl(\frac{a_s\lambda}{u_s}
+
a_m\Bigr)(1+8a^2_m)\Bigr] + \frac{16a_s\lambda^3}{u_s}\left[1+2\left(\frac{a_s\lambda}{u_s}+2a_s\right)\right]\\
\hspace{0.77cm}\left.-(37u^2_m+25g^2_m-10u_mg_m)
\Bigl[1+4\Bigl(\frac{a_s\lambda}{u_s}+a_m\Bigr)\Bigr]
+4608a^3_mu_m(2u_m-g_m)\Bigl[1+10\Bigl(\frac{a_s\lambda}{u_s}+a_m\Bigr)\Bigr]\right\},\\
\frac{dg_m}{dl} = g_m +
\frac{1}{2\pi^2}\Bigl\{\frac{32a_s\lambda^2}{u_s}(2g_m-u_m)\Bigl[(1+4a^2_m)(1+2a_s)
+4\Bigl(\frac{a_s\lambda}{u_s}+a_m\Bigr)(1+8a^2_m)\Bigr]\\
\hspace{0.77cm}-4a^2_m\Bigl[(g_m+73u_m)(7g_m-5u_m)+9(u_m-g_m)^2\Bigr]
\Bigl[1+8\Bigl(\frac{a_s\lambda}{u_s}+a_m\Bigr)\Bigr]\\
\hspace{0.77cm}-\Bigl[12(u_m-g_m)(2g_m-u_m)+6(u_m+g_m)^2-\frac{\lambda^2}{4}\Bigr]
\Bigl[4\Bigl(\frac{a_s\lambda}{u_s}+a_m\Bigr)+1\Bigr]\\
\hspace{0.77cm}+576a_mu_m(2g_m-u_m)\Bigl[6\Bigl(\frac{a_s\lambda}{u_s}+a_m\Bigr)+1\Bigr]\Bigr\},\\
\frac{du_s}{dl}
=u_s+\frac{4}{\pi^2}\left\{\frac{27a_su^2_s}{2}(1+6a_s)
-\frac{144a_sa_mu_m\lambda^3}{u_s}\Bigl[1+10\Bigl(\frac{a_s\lambda}{u_s}+a_m\Bigr)\Bigr]
-\frac{\lambda^2}{2}\Bigl[4\Bigl(\frac{a_s\lambda}{u_s}+a_m\Bigr)+1\Bigr]\right.\\
\hspace{0.77cm}-\frac{9u^2_s}{16}(4a_s+1) - \frac{4\lambda^2_{\Delta
A}}{3} \Bigl(\frac{4\lambda_{\Delta A}a_s}{u_s}+1\Bigr)
-a^2_m\lambda^2\Bigl[1+8\Bigl(\frac{a_s\lambda}{u_s}+a_m\Bigr)\Bigr]\\
\hspace{0.77cm}\left.+\frac{8a_s\lambda^3}{u_s}
\left(1+\frac{6a_s\lambda}{u_s}\right)+\frac{2768a_s\lambda^3_{\Delta
A}}{35u_s}
\left(1+\frac{6a_s\lambda_{\Delta A}}{u_s}\right)\right\},\\
\frac{d\lambda}{dl}
=\lambda+\frac{\lambda}{\pi^2}\left\{9a_su_s(6a_s+1)
+72a_mu_m\Bigl[6\Bigl(\frac{a_s\lambda}{u_s}+a_m\Bigr)+1\Bigr]
-(2u_m-g_m)\Bigl[4\Bigl(\frac{a_s\lambda}{u_s}+a_m\Bigr)+1\Bigr]\right.\\
\hspace{0.77cm}-\frac{3u_s}{8}(4a_s+1)+\frac{96a_sa^2_m\lambda^2}{u_s}
\Bigl(\frac{a_s\lambda}{u_s}+a_m\Bigr)-2\lambda\Bigl[1-\frac{2a_s\lambda}{u_s}(1+4a^2_m)\Bigr]
\Bigl[2a_s+2\Bigl(\frac{a_s\lambda}{u_s}+a_m\Bigr)+1\Bigr]\\
\hspace{0.77cm}-4a^2_m(20u_m-g_m)\Bigl[1+8\Bigl(\frac{a_s\lambda}{u_s}+a_m\Bigr)\Bigr]+576a^3_mu_m
\Bigl(1+10\Bigl(\frac{a_s\lambda}{u_s}+a_m\Bigr)\Bigr)\\
\hspace{0.77cm}\left.+\frac{16a_s\lambda}{u_s}(2u_m-g_m)\left(1+\frac{6a_s\lambda}{u_s}\right)
+6a_s\lambda\left[1+2a_s\left(\frac{\lambda}{u_s}+2\right)\right]\right\},\\
\frac{d\lambda_{\Delta A}}{dl} =\lambda_{\Delta A}+
\frac{\lambda_{\Delta A}}{3\pi^2}\left\{27a_su_s(1+6a_s)
-\frac{9u_s}{8}(4a_s+1)+\frac{64a_s\lambda^2_{\Delta A}}{u_s}
\Bigl[1+2a_s\Bigl(1+\frac{2\lambda_{\Delta A}}{u_s}\Bigr)\Bigr]\right.\\
\hspace{0.77cm}\left.-12\lambda_{\Delta
A}\Bigl[1+2a_s\Bigl(1+\frac{\lambda_{\Delta A}}
{u_s}\Bigr)\Bigr]+36a_s\lambda_{\Delta
A}\left[1+2a_s\left(2+\frac{\lambda_{\Delta A}}
{u_s}\right)\right]\right\}.
\end{array}\right.
\end{eqnarray}
These equations are used to analyze the impact of ordering
competition on the global phase diagram in the main context of the
paper.

\end{widetext}


\begin{thebibliography}{10}

\bibitem{LeeRMP2006}
P. A. Lee, N. Nagaosa, and X.-G. Wen, Rev. Mod. Phys. {\bf 78}, 17
(2006).

\bibitem{Loehneysen}
H. v. L\"{o}hneysen, A. Rosch, M. Vojta, and P. W\"{o}lfle, Rev.
Mod. Phys. {\bf 79}, 1015 (2007).

\bibitem{Stockert}
O. Stockert, S. Kirchner, F. Steglich, and Q. Si, J. Phys. Soc. Jpn.
{\bf 81}, 011001 (2012).

\bibitem{Kamihara2008JACP}
Y. Kamihara, T. Watanabe, M. Hirano, H. Hosono, J. Am. Chem. Soc.
{\bf 130}, 3296 (2008).

\bibitem{Chen2008Nature}
X. H. Chen, T. Wu, G. Wu, R. H. Liu, H. Chen, and D. F. Fang, Nature
(London) {\bf 453}, 761 (2008).

\bibitem{Chen2008PRL}
G. F. Chen, Z. Li, D. Wu, G. Li, W. Z. Hu, J. Dong, P. Zheng,
J. L. Luo, and N. L. Wang, Phys. Rev. Lett. {\bf 100}, 247002 (2008).

\bibitem{Rotter2008PRL}
M. Rotter, M. Tegel, and D. Johrendt, Phys. Rev. Lett. {\bf 101},
107006 (2008).

\bibitem{Fisher2011RPP}
I. R. Fisher, L. Degiorgi, and Z. X. Shen, Rep. Prog. Phys. {\bf
74}, 124506 (2011).

\bibitem{Hirschfeld2011RPP}
P. J. Hirschfeld, M. M. Korshunov, and I. I. Mazin, Rep. Prog. Phys.
{\bf 74}, 124508  (2011).

\bibitem{Chubukov2012}
A. V. Chubukov, Annu. Rev. Condens. Matter Phys. {\bf 3}, 57 (2012).

\bibitem{Vojta_rev}
M. Vojta, Adv. Phys. {\bf 58}, 699 (2009); E. Fradkin \emph{et al.},
Annu. Rev. Condens. Matter Phys. {\bf 1}, 153 (2010).

\bibitem{Liu2012PRB}
G.-Z. Liu, J.-R. Wang, and J. Wang, Phys. Rev. B {\bf 85}, 174525
(2012).

\bibitem{Hashimoto2012Science}
K. Hashimoto, K. Cho, T. Shibauchi, S. Kasahara, Y. Mizukami, R.
Katsumata, Y. Tsuruhara, T. Terashima, H. Ikeda, M.A. Tanatar, H.
Kitano, N. Salovich, R.W. Giannetta, P.Walmsley, A. Carrington, R.
Prozorov, and Y. Matsuda, Science {\bf 336}, 1554 (2012).

\bibitem{Levchenko2013PRL}
A. Levchenko, M. G. Vavilov, M. Khodas, and A.V. Chubukov, Phys.
Rev. Lett. {\bf 111}, 177003 (2013).

\bibitem{Chowdhury2013PRL}
D. Chowdhury, B. Swingle, E. Berg, and S. Sachdev, Phys. Rev. Lett.
{\bf 111}, 157004 (2013).

\bibitem{Fernandes2013PRL}
R. M. Fernandes, S. Maiti, P. W\"{o}lfle, and A. V. Chubukov,
Phys. Rev. Lett. {\bf 111}, 057001 (2013).

\bibitem{Wilson1975RMP}
K. G. Wilson, Rev. Mod. Phys. {\bf 47}, 773 (1975).

\bibitem{Shankar1994RMP}
R. Shankar, Rev. Mod. Phys. {\bf 66}, 129 (1994).

\bibitem{Moon2010PRB}
E. G. Moon and S. Sachdev, Phys. Rev. B {\bf 82}, 104516 (2010).

\bibitem{Fernandes2014NPhys}
R. M. Fernandes, A. V. Chubukov, and J. Schmalian, Nat. Phys. {\bf
10}, 97 (2014).

\bibitem{Nandi}
S. Nandi, M. G. Kim, A. Kreyssig, R. M. Fernandes, D. K. Pratt, A.
Thaler, N. Ni, S. L. Bud'ko, P. C. Canfield, J. Schmalian, R. J.
McQueeney, and A. I. Goldman, Phys. Rev. Lett. {\bf 104}, 057006
(2010).

\bibitem{Maiti2010PRB}
S. Maiti and A. V. Chubukov, Phys. Rev. B {\bf 82}, 214515 (2010).

\bibitem{Chubukov2008PRB}
A. V. Chubukov, D. V. Efremov, and I. Eremin, Phys. Rev. B {\bf 78},
134512 (2008)

\bibitem{Lee2009PRL}
F. Wang, H. Zhai, Y. Ran, A. Vishwanath, D.-H. Lee
Phys. Rev. Lett. {\bf 102}, 047005 (2009).

\bibitem{Medici2011PRB}
L. de' Medici, Phys. Rev. B {\bf 83}, 205112 (2011).

\bibitem{Schickling2012PRL}
T. Schickling, F. Gebhard, J. B\"{u}nemannand, L. Boeri, O. K.
Andersen, and W. Webe, Phys. Rev. Lett. {\bf 108}, 036406 (2012).

\bibitem{Medici2014PRL}
L. de' Medici, G. Giovannetti, and M. Capone, Phys. Rev. Lett. {\bf
112}, 177001 (2014).

\bibitem{Fernandes2012PRB}
R. M. Fernandes, A. V. Chubukov, J. Knolle, I. Eremin, and J.
Schmalian, Phys. Rev. B {\bf 85}, 024534 (2012).

\bibitem{She2010PRB}
J.-H. She, J. Zaanen, A. R. Bishop, and A. V. Balatsky, Phys. Rev. B
{\bf 82}, 165128 (2010).

\bibitem{Millis2010}
Z. Nussinov, I. Vekhter and A. V. Balatsky, Phys. Rev. B {\bf 79},
165122 (2009); A. J. Millis, Phys. Rev. B {\bf 81}, 035117 (2010).

\bibitem{Wang2014PRD}
J. Wang and G.-Z. Liu, Phys. Rev. D {\bf 90}, 125015 (2014).

\bibitem{Ikeda2013PRL}
T. Nomoto and H. Ikeda, Phys. Rev. Lett. {\bf 111}, 167001 (2013).

\bibitem{Suppl}
See Supplemental Material at http:.... for more detailed analytical
treatment of the effective theory and for the expression of the
coupled RG equations of all the effective parameters.

\bibitem{Halperin1974PRL}
B. I. Halperin, T. C. Lubensky, and S.-K. Ma, Phys. Rev. Lett. {\bf
32}, 292 (1974).

\bibitem{Schmalian2010PRB}
R. M. Fernandes and J. Schmalian, Phys. Rev. B {\bf 82}, 014521 (2010).

\bibitem{Domany_Chen_Rudnick_Iacobson}
E. Domany, D. Mukamel, and M. E. Fisher, Phys. Rev. B {\bf 15}, 5432
(1977); J. H. Chen, T. C. Lubensky, and D. R. Nelson, Phys. Rev. B
{\bf 17}, 4274 (1978); J. Rudnick, Phys. Rev. B {\bf 18}, 1406
(1978); H. H. Iacobson and D. J. Amit, Ann. Phys. {\bf 133}, 57
(1981).

\bibitem{Wang2014PRB}
J. Wang, A. Eberlein, and W. Metzner, Phys. Rev. B {\bf 89},
121116(R) (2014).

\bibitem{Pratt2009PRL}
D. K. Pratt, W. Tian, A. Kreyssig, J. L. Zarestky, S. Nandi, N. Ni,
S. L. Bud'ko, P. C. Canfield, A. I. Goldman, and R. J. McQueeney,
Phys. Rev. Lett. {\bf 103}, 087001 (2009).

\bibitem{Christianson2009PRL}
A. D. Christianson, M. D. Lumsden, S. E. Nagler, G. J. Mac-Dougall,
M. A. McGuire, A. S. Sefat, R. Jin, B. C. Sales, and D. Mandrus,
Phys. Rev. Lett. {\bf 103}, 087002 (2009).

\bibitem{Kreyssig2010PRB}
A. Kreyssig, M. G. Kim, S. Nandi, D. K. Pratt, W. Tian, J. L.
Zarestky, N. Ni, A. Thaler, S. L. Bud'ko, P. C. Canfield, R. J.
McQueeney, and A. I. Goldman, Phys. Rev. B {\bf 81}, 134512 (2010).

\bibitem{Fernandes1504.03656}
R. M. Fernandes, S. A. Kivelson, and E. Berg, arXiv:1504.03656.

\bibitem{She2015PRB}
J.-H. She, M. J. Lawler, and E.-A. Kim, Phys. Rev. B {\bf 92}, 035112 (2015).

\bibitem{Hashimoto2012PNAS}
K. Hashimoto, Y. Mizukami, R. Katsumata, H. Shishido, M. Yamashita,
H. Ikeda, Y. Matsuda, J. A. Schlueter, J. D. Fletcher, A.
Carrington, D. Gnida, D. Kaczorowski, and T. Shibauchi, PNAS {\bf
110}, 3293 (2013).

\end{thebibliography}
\end{document}